# The role of precursor coverage in the synthesis and substrate transfer of graphene nanoribbons

*Rimah Darawish[1,2], Oliver Braun[3,4], Klaus Müllen[5,6], Michel Calame[3,4,7], Pascal Ruffieux[1], Roman Fasel[1,2] and Gabriela Borin Barin[1]\**

[1]Empa, Swiss Federal Laboratories for Materials Science and Technology, nanotech@surfaces Laboratory, 8600 Dübendorf, Switzerland

[2]Department of Chemistry, Biochemistry and Pharmaceutical Sciences, University of Bern, 3012 Bern, Switzerland

[3]Empa, Swiss Federal Laboratories for Materials Science and Technology, Transport at Nanoscale Interfaces Laboratory, 8600 Dübendorf, Switzerland

[4]Department of Physics, University of Basel, 4056 Basel, Switzerland

[5]Max Planck Institute for Polymer Research, 55128 Mainz, Germany

[6]Department of Chemistry, Johannes Gutenberg University Mainz, Duesbergweg 10-14, 55128, Mainz, Germany

[7]Swiss Nanoscience Institute, University of Basel, 4056 Basel, Switzerland

\*corresponding author: gabriela.borin-barin@empa.ch

**Abstract**

Graphene nanoribbons (GNRs) with atomically precise widths and edge topologies have well-defined band gaps that depend on ribbon dimensions, making them ideal for room-temperature switching applications like field-effect transistors (FETs). For the effective integration of GNRs into devices, it is crucial to optimize growth conditions to maximize GNR length and, consequently, device yield. Additionally, it is essential to establish device integration and monitoring strategies that maintain GNR quality during the transition from growth to device fabrication. In this study, we investigate the growth and alignment of 9-atom-wide armchair graphene nanoribbons (9-AGNRs) on a vicinal gold substrate, Au(788), with varying molecular precursor doses (PD) and, therefore, different resulting GNR coverages. Our investigation reveals that the GNR growth location on the Au(788) substrate is coverage-dependent. Furthermore, scanning tunneling microscopy shows a strong correlation between the GNR length evolution and both the PD and the GNR growth location on the substrate. Employing Raman spectroscopy, we analyze samples with eight different PDs on the growth substrate. We find that GNR alignment improves with length, achieving near-perfect alignment with an average GNR length of ~40 nm for GNRs growing solely at the Au(788) step edges. To fully exploit GNR properties in device architectures, GNRs need to be transferred from their metallic growth substrate to semiconducting or insulating substrates, such as $SiO_2$/Si or $Al_2O_3$. Upon substrate transfer, samples produced with higher PD present systematically better alignment preservation and less surface disorder, which we attribute to reduced GNR mobility during the transfer process. Importantly, we observe that PD also affects the substrate transfer success rate, with higher success rates observed for samples with higher GNR coverages (77%) compared to those with lower GNR coverages (53%). Our findings characterize the important relationship between precursor dose, GNR length, alignment quality, and surface disorder during GNR growth and upon substrate transfer, offering crucial insights for the further development of GNR-based nanoelectronic devices.

**Introduction**

Atomically precise graphene nanoribbons (GNRs) have gathered significant interest in recent years due to their tunable physicochemical properties[1–3], achieved through precise control over their width[4–7] and edge structure[8–12]. This makes GNRs appealing candidates for various electronic[13–24], spintronic [25–27], and optical applications[2,3,28,29].

Among the different types of GNRs, armchair-edged graphene nanoribbons (AGNRs) have attracted particular attention due to their width-dependent electronic band gap, which can be adjusted from quasi-metallic to wide band gap semiconductors[30]. AGNRs are classified into three families based on their width: $N=3p$ (medium gap), $N=3p + 1$ (wide gap), and $N=3p + 2$ (quasi-metallic / narrow gap), where $p$ is an integer and $N$ indicates the number of carbon atoms along the GNR width[30]. However, achieving a well-defined band gap requires precise control over the ribbons' width, edge structure and length.

To synthesize atomically precise GNRs, on-surface synthesis is a versatile approach that involves assembling molecular building blocks on a catalyst substrate (usually Au(111)) under ultrahigh vacuum conditions[7,10]. This process is based on depositing suitably designed molecular precursors on the metal surface, followed by their surface-assisted covalent coupling. By carefully designing the precursor monomer, atomic precision over GNR width and edge topology is achieved, enabling the synthesis of ultra-narrow GNRs with atomically precise widths (5-[5], 7-[7], 9-[4], 13-[31] and 17[32]-AGNRs) and defined edge topology (armchair-[4], zigzag-[10], chiral-[33], and GNRs with topological phases[12]).

To fully exploit GNR properties in device architectures, GNRs need to be transferred from their metallic growth substrate to semiconducting or insulating substrates, such as $SiO_2$/Si[34,35]. Various methods have been developed to transfer GNRs, depending on whether the growth substrate is a gold film or a single crystal. A polymer-free transfer is typically used for GNRs grown on Au(111) films on mica, using the gold film itself to support the GNRs throughout the transfer[35]. In the case of GNRs growing uniaxially aligned on a regularly stepped gold single crystal surface such as Au(788), the method of choice is an electrochemical delamination transfer, primarily developed for graphene transfer from copper foils[36], and later optimized for GNRs[29,34]. This approach uses a polymer layer, usually poly(methyl methacrylate) (PMMA), to support the GNRs, and relies on water electrolysis to generate hydrogen bubbles at the interface between the PMMA/GNRs layer and the metal substrate. The hydrogen bubbles mechanically delaminate the PMMA/GNRs layer from the metal substrate, resulting in the transfer of uniaxially aligned GNRs[29,34,37].

Among AGNRs, 9-atom-wide armchair GNRs (9-AGNRs) have been most extensively integrated into devices due to their suitable electronic gap (1.4 eV measured on Au)[4] enabling switching behavior at room temperature[38], suitable length to bridge source and drain contacts[39] and mechanical robustness and chemical stability under ambient conditions[35]. Another important aspect of the integration of AGNRs into devices is the device yield, which typically ranges between 10-15% when using AGNRs grown on Au (111) surfaces, due to their non-preferential growth direction[40]. By growing uniaxially aligned AGNRs, device yields can reach ~85%, as the GNRs can be deposited aligned with the source to drain direction of the device[20].

In this work, we investigate the growth of aligned 9-AGNRs on Au(788) and characterize their length as a function of precursor dose (PD) in 30 different samples using scanning tunneling microscopy (STM). We also characterize the quality of alignment and surface disorder for the different sample as a function of PD and length on both the growth substrate (40 samples) and after substrate transfer (27 samples) using polarized Raman spectroscopy. Our results demonstrate that PD plays a key role in determining the quality of alignment and surface disorder, and significantly impacts the success rate of GNRs' substrate transfer.

**Results and Discussion**

*9-AGNR growth and length evolution on Au (788)*

To investigate the growth of 9-AGNRs on a vicinal surface, we deposit the precursor molecule 3′,6′-di-iodine-1,1′:2′,1″-terphenyl (DITP)[39] onto Au(788) (kept at room temperature) at a fixed deposition rate of 1 Å/min (as measured with a quartz microbalance), with deposition times varying from 1 to 9 minutes. Subsequently, a 2-step process thermal annealing at 200°C and 400°C activate polymerization and induce cyclodehydrogenation, respectively, to form the final GNR structure[4]. By maintaining a constant deposition rate of 1 Å/min and increasing the deposition time in 1-minute steps, we achieve PDs ranging from 1 to 9 Å on the surface. Figs. 1a-h shows representative STM images of 9-AGNRs samples with PDs of 1 to 8-9 Å (8 or 9 Å = 1 full monolayer), respectively.

We observe that, as PD increases, the growth of 9-AGNRs on Au(788) occurs at three different positions. Initially, GNRs start growing along the Au(788) step edges, referred here as the 1$^{st}$-row position (blue arrow). A representative STM image of 9-AGNRs with PD = 1 Å is presented in Fig. 1a, with short GNRs with an average length of 14 nm, growing solely at this position. As PD increases to 2 and 3 Å, GNRs continue to grow exclusively at the step edges with average GNR length reaching 19 nm and 35 nm for PDs of 2 Å and 3 Å, respectively (Figs. 1b-c). The growth of GNRs at the 1$^{st}$-row position only, along the step edges, can be attributed to the higher catalytic activity and altered surface chemistry caused by the greater negative charge density at the lower step edge[41,42], which facilitates the nucleation and growth process. For samples with low PD between 1 and 3 Å, the inter-ribbon distance remains constant and is determined by the width of the terraces, approximately 3.8 nm[43].

For samples prepared with PD of 4 Å, the lower step edges are fully decorated with 9-AGNRs, and additional GNRs start to grow in the middle of the Au(788) terraces, referred here as the 2$^{nd}$-row position (Fig. 1d, green arrow). Initially, GNRs at this position are also short with an average length of 14 nm, due to the low amount of precursors available on the terraces. In some instances, such short GNRs also grow misaligned to the step edge (Fig. 1d, black arrow).

As PD further increases to 5 Å, 6 Å, and 7 Å, the average length of GNRs at the 2$^{nd}$-row position steadily increases to 22 nm, 30 nm, and 36 nm, respectively (Figs. 1e-g). Notably, at PD = 7 Å, short GNRs (red arrow, Fig. 1g) also grow at the 3$^{rd}$-row position close to the descending step edge of the terrace, with an average length of 13 nm. At PD = 8-9 Å, a complete monolayer is formed, with three rows of ribbons per terrace

(Fig. 1h). In the monolayer samples, GNRs at the 1st-row exhibit an average length of 46 nm, the 2nd-row GNRs of 42 nm, and the 3rd-row GNRs of 36 nm.

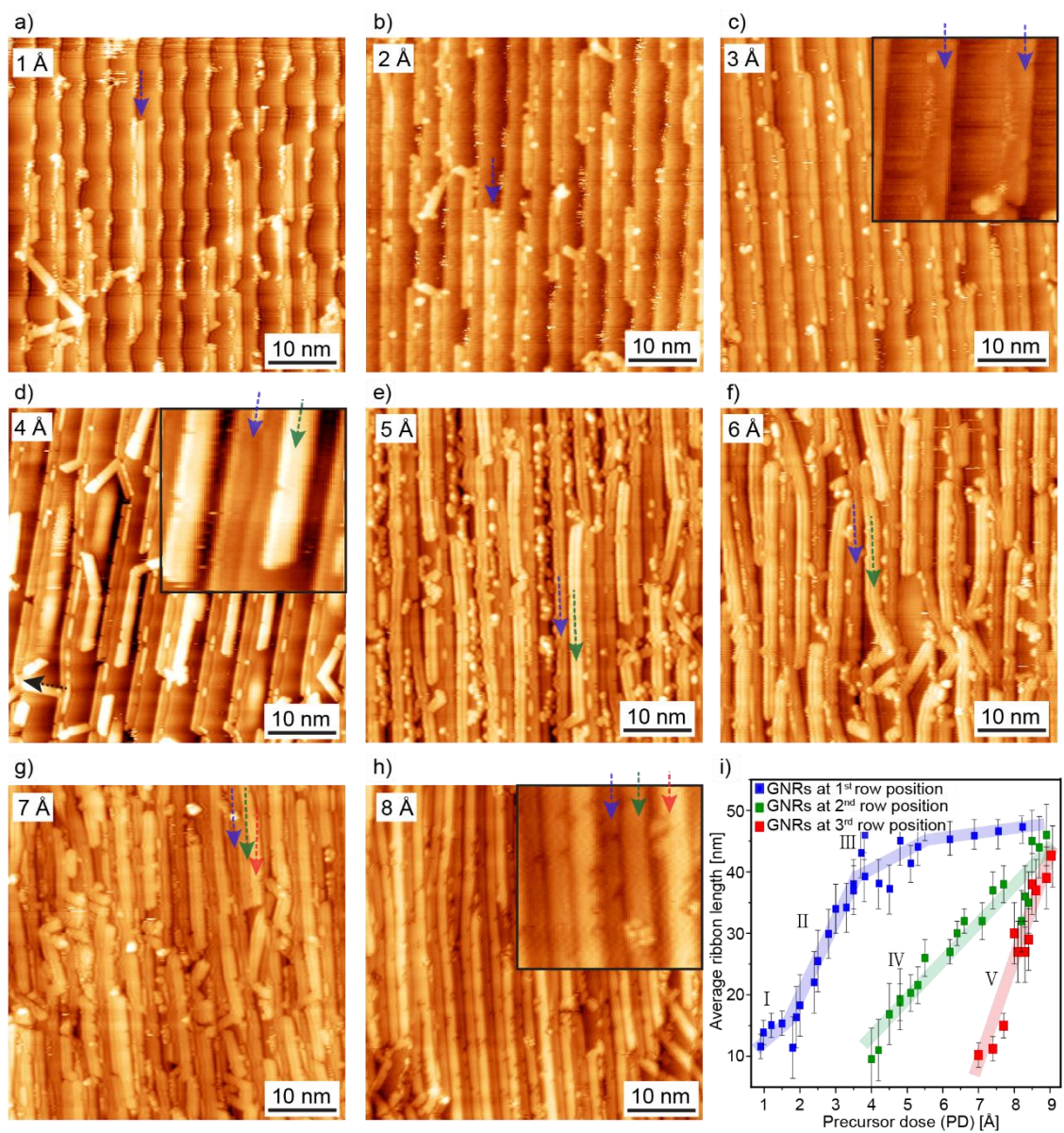

**Figure 1: Evolution of 9-AGNR growth on Au(788) with increasing precursor dose (PD).** (a-c) STM images with GNR length evolution at the 1st-row position (blue arrows) along the lower step edges as PD increases from 1 Å to 3 Å. d) STM image of GNRs at the 2nd-row position (green arrow) growing parallel to 1st-row GNRs (blue) at PD = 4 Å. (e-f) STM images revealing the length evolution of GNRs at the 2nd-row position (green arrows) at PD = 5 Å and PD = 6 Å, respectively. g) STM image at PD = 7 Å, revealing the formation of a GNR at the 3rd-row position (red arrow), growing parallel to the GNRs grown previously at the 1st- and 2nd-row positions. h) STM image of a monolayer of 9-AGNRs with GNRs grown in three parallel

rows (blue, green, and red arrows) along each terrace at PD = 8 Å. i) Histogram illustrating GNR length evolution for 30 samples across all PDs, with error bars representing standard deviation. Tunneling parameters for STM images: $V_b$ = -1.5 V, $I_t$ = 30 pA. Insets in c), d) and h) are zoom-in of the same images for better visualization.

Figure 1i shows an overview of the length evolution of GNRs, displaying the average length distribution based on measurements of a total of 30 samples for PD ranging from 1 to 9Å. Large-scale STM images (100 x 100 nm) were acquired, allowing for the examination of more than 800 GNRs per sample. As a complement, the average GNR lengths for selected PDs is summarized in Table S1. The length evolution data displayed in Fig. 1i highlight the significantly different growth rates for GNRs growing at $1^{st}$-, $2^{nd}$-, and $3^{rd}$-row positions, as well as a non-linear growth behavior for $1^{st}$-row GNRs, which, however, can be approximated by a sequence of three linear regimes. We performed linear curve fitting to the data in Fig. 1i for each of the 5 regimes I to V of GNR length evolution, as discussed in the following.

Regime I corresponds to GNRs growing with PDs ranging from 0 to 2 Å. At these very low precursor coverages, GNR growth is dominated by nucleation at the step edges, more specifically at defect sites. This leads to a low growth rate of 5 nm Å$^{-1}$, and the average length of the nucleating GNRs thus only increases by 5 nm as PD increases from 1 to 2 Å. Upon increasing PD beyond 2 Å (regime II), the growth rate increases significantly to 16 nm Å$^{-1}$, with the average GNR length reaching 35 nm at PD = 3 Å. This corresponds to the situation where further GNR nucleation at the step edge is low, and most incoming precursors contribute to increasing the length of GNRs growing along the step edges. In regime III, starting at PD = 4 Å, all step edges become saturated, with only an incremental increase (4 nm Å$^{-1}$) in the average length of GNRs, up to the final average $1^{st}$-row GNR length of 43 nm. Before the step edges are fully passivated, GNRs also start to grow at the $2^{nd}$-row position (on the Au (788) terraces), here defined as regime IV (PD = 4 to 9 Å). In this regime a strictly linear length evolution occurs, leading to a steady increase in GNR length by 8 nm Å$^{-1}$. We attribute this relatively slow growth rate to a combination of significant nucleation density and the fact that, in this regime of $2^{nd}$-row GNR growth, incoming precursors also contribute to complete $1^{st}$-row GNRs and nucleation/lengthening of $3^{rd}$-row GNRs. Finally, when the $2^{nd}$-row

GNR growth approaches saturation, GNRs start to grow a the 3$^{rd}$-row position, close to the upper step edge of each terrace. This is regime V, with a high growth rate of 23 nm Å$^{-1}$, leading to a 3$^{rd}$-row GNR length increase from 13 to 36 nm as PD increases from 7 to 9 Å. We attribute this rapid increase in GNR length to the presence of fewer nucleation sites, therefore the incoming precursors mostly contribute to the elongation of GNRs at this position.

As here, it appears that GNR growth at 1$^{st}$-, 2$^{nd}$-, and 3$^{rd}$-row positions is not a strictly sequential process, but Fig. 1i clearly shows that growth in the next row position starts before the previous row has been fully completed, as evidenced by the overlap of regimes III and IV as well as IV and V. This simply reflects the balance between nucleation and diffusion at play, which may be somewhat influenced by the growth temperature. We have not explored this aspect, which thus remains for future work. Overall, the detailed findings discussed above provide important insight into the growth of GNRs on vicinal surfaces and highlight the influence of growth position on the substrate and PD on GNR length evolution.

**9-AGNR quality and alignment as a function of PD on Au(788) and after substrate transfer**

After transferring GNRs onto an insulating substrate, characterization using STM becomes infeasible due to the requirement for a conductive sample surface. Raman spectroscopy provides a non-destructive and rapid alternative for assessing GNR quality and orientation regardless of the substrate's nature.[29,34,44] Thus, a detailed Raman analysis for the samples studied above was performed.

The main Raman active mode for GNRs is the G mode, located in the high-frequency spectral range at ~1600 cm$^{-1}$ [45–47,35]. This mode is present in all sp$^2$ carbon-based materials and originates from in-plane vibrations[48]. Additionally, several phonon modes are detected in the high-frequency spectral range of 1100-1500 cm$^{-1}$, which are associated with the edge structure of GNRs and collectively referred to as CH-D modes[35,45–47]. Specifically, the CH-bending mode at ~1200 cm$^{-1}$ corresponds to the bending vibrations of the hydrogen atoms at the GNR edges, while the D mode at ~1300

cm$^{-1}$ indicates disarrangement of the periodic graphene honeycomb lattice. It is important to note that unlike in graphene, the D peak in GNRs is an intrinsic mode resulting from the precise atomic edges[35], rather than defects[49]. Another intrinsic peak in GNRs is the radial breathing-like mode (RBLM), which is located in the low-frequency range and provides information about GNR width[35,50,51]. This mode is similar to the radial breathing mode (RBM) in carbon nanotubes (CNTs), whose frequency is directly related to the diameter and chirality of the CNTs[52].

Here, Raman measurements are conducted in a home-built vacuum chamber (~10$^{-2}$ mbar) using 785 nm excitation wavelength (1.58 eV). In addition, an optimized mapping approach (maps of 10 μm x 10 μm, 10 x 10 pixels) is used to further limit damage to the GNRs[34]. Figure 2a shows representative Raman spectra acquired directly on the growth substrate Au(788) for 9-AGNRs with PDs from 1 to 8 Å. In all spectra, the RBLM, CH, D, and G modes are observed, with the intensity of the Raman peaks increasing proportionally with PD. In samples with high PD, we observe an additional peak in the CH-D area at ~1302 cm$^{-1}$ on Au(788) and at ~1285 cm$^{-1}$ after substrate transfer. This additional peak could be related to the interaction between GNRs, likely due to the high coverage and shorter inter-GNR distances (Figs. 2a and 2b).

To investigate the electronic properties of GNRs in a device architecture, a substrate transfer process is required. Here, aligned 9-AGNRs are transferred from Au(788) onto a Raman-optimized substrate (ROS)[34] using electrochemical delamination transfer[29,34]. Figure 2b depicts Raman spectra of 9-AGNRs with PD from 1-8 Å after substrate transfer onto a ROS. The presence of all 9-AGNR intrinsic Raman peaks (RBLM, CH, D, and G) after the substrate transfer process suggests the preservation of the GNRs' structural integrity during transfer. Aditionally, the presence of additional modes are also observed, which we attribute to overtones, including the RBLM3 at 845 cm$^{-1}$ [34].

To evaluate GNR quality before and after substrate transfer, peak positions and the full-width at half maximum (FWHM) of RBLM, CH, D, and G modes are extracted for all samples. The results are summarized in Table S2 and Table S3, respectively.

Upon substrate transfer, samples with medium and high PDs (4-8Å) exhibit negligible changes in the FWHM for RBLM, D, CH, and G Raman modes, with all values falling within the experimental accuracy of the Raman measurements (3 cm$^{-1}$). However, low-PD samples (1-3 Å), present a significant standard deviation in the FWHM between the samples in this range, particularly for the CH mode. The CH mode is an edge mode, and the increase in FWHM could be attributed to the presence of defects at the edges induced by the transfer process. This effect is particularly prominent in low PD samples where GNRs are strongly attached to the step edges, leading to lower GNR quality upon transfer[37]. Evaluation of the peak positions of the Raman active modes for all samples reveals negligible shifts in the G, D, and RBLM modes. However, the CH vibrational modes exhibit a notable downshift of 10-12 cm$^{-1}$ across all PD samples, which may be attributed to the different interaction of the edge modes with the ROS substrate compared to Au.

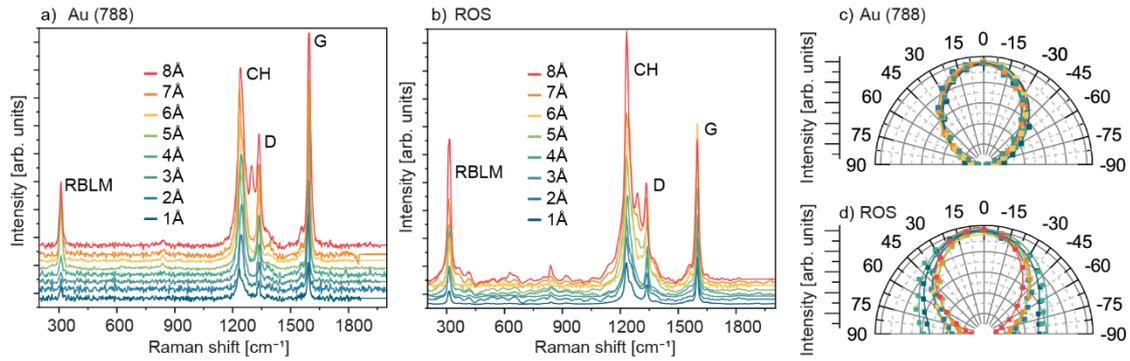

**Figure 2. Raman characterization of 9-AGNRs** at PDs from 1-8 Å on Au(788) (a,c) and after substrate transfer onto ROS (b,d). Panels. (c) and (d) polar plots of the polarization dependence of the G-mode for each PD on Au(788) and the ROS, respectively. The intensities in the polar plots are normalized to the 8 Å precursor dose. The incident light is polarized at an angle $\theta_{in}$ with respect to the orientation of the gold terraces (GNRs' alignment direction ($\theta = 0°$)), and the backscattered light is detected without an analyzing polarizer. The polar diagrams follow the $cos^2(\theta)$ function as expected for this configuration and are fitted with a modified angular model based on Ref [37] (lines). The Raman measurements are obtained under vacuum conditions with a 785 nm excitation wavelength.

To achieve optimal performance in GNR-based FET devices and improve device yield[20], it is crucial to determine the alignment direction of GNRs such as to transfer them well aligned with the source to drain direction. Here we use polarized Raman spectroscopy to investigate the orientation of GNRs on the growth substrate and after substrate transfer[29,37]. The laser excitation source (785 nm) is polarized from -90° to 90° at 10° increments, and the scattered light is detected without using an analyzing polarizer to maximize the overall signal (the spectrograph collects scattering data for all polarizations).

By measuring the polarization-dependent intensity of scattered light, we observe that the Raman intensity of the G mode follows a $\cos^2(\theta_{in})$ polarization-dependent behavior, with maximum intensity at 0° and minimum at 90°, as shown in the polar plots in Figs. 2c and 2d. To quantitatively characterize the alignment of GNRs, two methods are employed. The first is the Raman polarization anisotropy ($P$) [29,37,44], defined by Eq. (1):

$$P = (I_\parallel - I_\perp) / (I_\parallel + I_\perp) \qquad (1)$$

where $I_\parallel$ and $I_\perp$ represent the Raman intensities with polarization along and perpendicular to the GNR axis, respectively. A perfect uniaxial alignment of GNRs corresponds to $P=1$, while $P=0$ indicates random orientation. The second approach is an extended Gaussian distribution, initially detailed in our earlier work[37]. Here our previous model is modified, to take into account that no analyzing polarizer was used during the measurements, as in Eq. (2).

$$I_{exp}(\vartheta) = A \cdot \int_{0°}^{360°} \cos^2(\varphi) \cdot \frac{1}{\sigma\sqrt{2\pi}} e^{-\frac{(\vartheta-\varphi-\vartheta_0)^2}{2\sigma^2}} d\varphi + B \cdot \frac{1}{2} \qquad (2)$$

Where $\vartheta_0$ is the azimuthal angle along which GNRs are preferentially aligned; $A$ is the fraction of surface area that exhibits aligned GNRs; $\sigma$ is the width of the Gaussian distribution characterizing the angular distribution around $\vartheta_0$ (quality of alignment); and $B$ is the fraction of the surface area contributing to the isotropic, polarization-independent Raman signal. From $A$ and $B$ the overall disorder present on the surface is defined as: $OD=B/(B+A)\cdot100\%$[37].

Here, both methods are applied to investigate the alignment of GNRs for all PDs. To guarantee representative results 40 samples with PD ranging between 1-9 Å on the Au (788) growth surface and 27 samples after substrate transfer onto ROS are investigated. We extract $P$, $\sigma$, and $OD$ for all Raman modes (RBLM, CH, D, and G), and summarize the results in Fig. 3 and Table 1 (for the G mode) and Fig. S1 (for the RBLM, CH, and D modes).

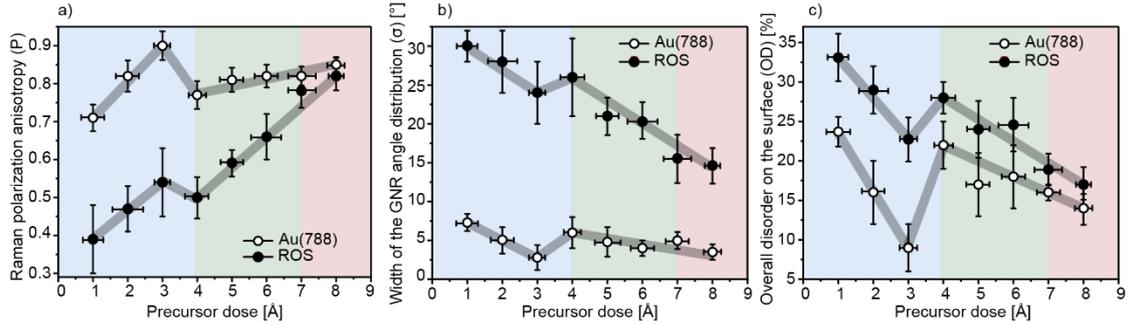

**Figure 3: Impact of PD on the alignment of GNRs on Au(788) and after substrate transfer onto ROS.** Panels (a) Polarization anisotropy (P), (b) width of GNR angle distribution ($\sigma$), and (c) overall disorder (*OD*) extracted for the G mode as a function of PD on both the Au(788) (open circles) and on ROS (closed circles) with error bars representing the standard deviation of the average values for each specific PD. The colored background represents the regimes in which GNRs grow preferentially at the 1$^{st}$-row position (blue), at the 2$^{nd}$-row position (green), and at the 3$^{rd}$-row position (red).

**Table 1:** Average *P*, $\sigma$, and *OD* of the G mode as a function of PD, on Au(788) and after substrate transfer to ROS.

| Parameters | Raman polarization anisotropy (*P*) | | Quality of alignment ($\sigma$) [°] | | Overall disorder (*OD*) [%] | |
|---|---|---|---|---|---|---|
| Substrate | Au(788) | ROS | Au(788) | ROS | Au(788) | ROS |
| 1 Å | 0.71± 0.04 | 0.39± 0.09 | 7± 1 | 30± 2 | 24± 2 | 33± 3 |
| 2 Å | 0.82± 0.04 | 0.47± 0.06 | 5± 2 | 28± 4 | 16± 4 | 29± 3 |
| 3 Å | 0.90 ± 0.04 | 0.55± 0.09 | 3± 1 | 24± 4 | 9± 4 | 23± 3 |
| 4 Å | 0.77± 0.02 | 0.50± 0.02 | 6± 2 | 26± 5 | 22± 3 | 28± 2 |
| 5 Å | 0.81± 0.03 | 0.65± 0.04 | 5± 2 | 21± 1 | 17± 4 | 24± 4 |
| 6 Å | 0.82± 0.03 | 0.70± 0.06 | 4± 1 | 20± 2 | 18± 4 | 25± 3 |
| 7 Å | 0.82± 0.03 | 0.78± 0.02 | 5± 1 | 16± 4 | 16± 1 | 19± 2 |
| 8-9 Å | 0.85± 0.02 | 0.82± 0.05 | 4± 1 | 15± 4 | 14± 3 | 17± 3 |

We start by discussing the low-PD regime, where GNRs grow solely on the step edges (highlighted in blue). In this regime, a clear trend is observed: $P_{Au(788)}$ increases and $\sigma_{Au(788)}$ becomes narrower as the length of GNRs increases, indicating a higher quality of alignment for longer GNRs on both Au(788) and ROS. GNRs grown at the 1$^{st}$-row position with average lengths of 14, 19, and 35 nm (for PD = 1, 2, and 3 Å, respectively) have $P_{Au(788)}$ = 0.71, 0.82, and 0.90, and $\sigma_{Au(788)}$= 7°, 5°, and 3°, respectively (Figs. 3a, 3b and Table 1). These results are consistent with our STM studies (Fig. 1) and reflect that longer GNRs tend to better align with the step edge, while shorter GNRs can also grow across the gold terraces. At PD = 4 Å, where GNR growth at the 2$^{nd}$-row position has started, we observe a kink in all curves of Figure 3. The overall degree of GNR alignment decreases, with $P_{Au(788)}$ = 0.77 and $\sigma_{Au(788)}$= 6° (Figs. 3a and 3b). For this PD, the step edges are fully saturated, and GNRs grow at the 2$^{nd}$-row position, located in the center of the terraces (highlighted in green), with an average length of 14 nm (Fig. 1d). As the 2$^{nd}$-row GNRs grow longer, the quality of alignment improves again, reaching $P_{Au(788)}$ = 0.81 and 0.82 with $\sigma_{Au(788)}$= 5 and 4° for precursor doses of 5 Å (average GNR length = 22 nm) and 6 Å (average GNR length = 30 nm), respectively. Finally, in the high-PD regime (highlighted in red), the substrate is fully covered with GNRs, and the quality of alignment reflects the behavior of GNRs grown at all three positions, as a full monolayer: with $P_{Au(788)}$ = 0.82 and 0.85, and $\sigma_{Au(788)}$= 5° and 4° for PD = 7 Å and 8-9 Å, respectively (Figs. 3a and 3b). The appearance of GNRs at the 3$^{rd}$-row position with an average length of 12 nm for PD = 7 Å does not negatively influence the quality of alignment, unlike the case of short GNRs grown with PD =1 Å and 4 Å (Fig. 2c). At this particular PD, GNRs at the 3$^{rd}$-row position grow well aligned because they are strictly confined between the 2$^{nd}$-row GNRs and the upper step edge of the adjacent substrate terrace.

The evolution of $P$, $\sigma$, with PD for GNRs transferred onto the ROS follows the same trend, with overall increasing quality of alignment with increasing coverage, except for the same kink at PD = 4 Å. However, compared to GNRs on Au(788), substrate-transferred GNRs exhibit a significantly decreased $P$ and an increased $\sigma$. In the low-PD regime (highlighted in blue), we extract $P_{ROS}$ = 0.39, 0.47, and 0.55, and $\sigma_{ROS}$ = of 30°, 28°, and 24°, for PD = 1, 2, and 3 Å respectively. We attribute the striking reduction in alignment after substrate transfer to the strong interaction of GNRs with the Au(788) step edges, which hinders their transfer, along with increased GNR mobility due to the low coverage preparation (Figs. 3a and 3b and polar plots in Fig. 2d). Longer GNRs at higher coverages, however, preserve their alignment better upon substrate transfer (Table 1). For samples with PD = 4 Å, the degree of alignment decreases slightly, with

$P_{ROS}$ = 0.5 and $\sigma_{ROS}$ = 26°, which we attribute to the presence of short GNRs growing at the 2$^{nd}$-row position (Figs. 3a and 3b and polar plots in Fig. 2d). These short 2$^{nd}$-row GNRs exhibit high mobility upon transfer, which negatively impacts their degree of alignment.

As PD increases, there is a steady increase in the preservation of alignment with $P_{ROS}$ = 0.65, 0.7, 0.78, 0.82, and $\sigma_{ROS}$ = 21°, 20°, 16°, and 15° for PD = 5 Å, 6 Å, 7 Å, and 8-9 Å, respectively. As the amount of GNRs on the surface increases (along with their length), their mobility during substrate transfer decreases, leading to a better preservation of alignment. Similar behavior was observed for full monolayer samples by Senkovskiy et al.[29], Overbeck et al.[44], Zhao et al.[2], and by us in our previous work[37], showing $P$ = 0.72-0.82 after transfer. Overbeck et al.[44] also reported the influence of GNR length on polarization dependence for short 5-AGNRs with an average length of 3.8 nm, where the low polarization anisotropy ($P_{Au(788)}$ = 0.3 before and $P_{ROS}$ = 0.4 after substrate transfer) was attributed to the GNRs' reduced shape anisotropy and the high mobility of short ribbons.

Next, we briefly discuss the overall disorder present on the surface (OD) for all samples before and after substrate transfer. The average values of OD for the G mode are presented in Fig. 3c and Table 1 (see Fig. S1 for the results for RBLM, CH, and D modes). For samples on Au(788), the origin of $OD_{Au(788)}$ is attributed to very short GNRs, irregularly fused precursor monomers, and the presence of impurities from the precursor[37]. Here, there is a systematic decrease of $OD_{Au(788)}$ as the GNRs grow longer: $OD_{Au(788)}$ = 24%, 16%, and 9%, for samples with PD = 1, 2, and 3 Å, respectively (Fig. 3c, highlighted in blue). As GNRs start to grow at the 2$^{nd}$-row position (PD = 4 Å), $OD_{Au(788)}$ increases to 22%, which reflects the presence of very short GNRs (that do not exhibit polarization dependence) growing on the terraces. As PD continues to increase (and therefore GNR length), $OD_{Au(788)}$ linearly decreases, reaching 14% for the full monolayer. Upon substrate transfer, $OD_{ROS}$ increases for all samples. However, the most significant increase is observed for low-PD samples, with $OD_{ROS}$ = 33%, 29%, and 23% (for PD = 1, 2, and 3 Å, respectively), compared to the full monolayer $OD_{ROS}$ = 17% (PD = 8-9 Å). This arises from an inefficient transfer in the case of low-PD samples, which may result in partially broken GNRs due to their strong physical interaction with the Au(788) step edges. Additionally, at these low precursor doses, the Au substrate is more exposed to PMMA and other contaminants, which may react with the Au surface and transfer along the GNRs, contributing to an increase in $OD_{ROS}$.

Finally, we comment on the success rate of transferring GNR samples grown from different PDs. Initially, 40 samples of 9-AGNR on the Au(788) substrate were fabricated and studied, of which 27 samples were successfully transferred to the ROS. The transfer success rate is larger for samples with higher PD, 77% for PD = 7-9 Å, compared to those with medium PD = 4-6 Å (60%) and low PD = 1-3 Å (53%), endorsing our results that GNRs at higher coverages transfer more efficiently.

**Conclusions**

We investigated the growth and alignment of 9-AGNRs on Au(788) with varying precursor doses, both on the growth substrate and upon substrate transfer. 40 samples with PDs ranging from 1 to 9 Å where characterized by STM. We observed that GNRs grow sequentially at three positions on Au(788) depending on PD: only at the lower step edges at low precursor doses (and thus low coverages), in the middle of the terraces at medium doses, and at the upper step edges at coverages approaching a complete monolayer. In terms of alignment, longer GNRs show better unidirectional alignment on Au(788), achieving near-perfect alignment for PD = 3 Å, when a single row of GNRs saturates the Au(788) step edges. As PD increases and GNRs start to grow in the center of the terraces the overall degree of alignment decreases due to the presence of short GNRs. A high degree of alignment is again obtained once a high PD is used to grow full monolayer samples. Upon substrate transfer, there is a significant decrease in the degree of alignment for low-PD samples, while alignment is largely preserved for high-PD ones. The overall disorder on the surface was also quantified. The presence of short GNRs increases the overall disorder on the growth surface, while a combination of short GNRs and impurities increases the overall disorder after substrate transfer. Finally, we also find a PD-dependent substrate transfer success rate, with samples grown from higher PD being more successfully transferred (77%) than lower PD samples (53%). Our work unravels the role of precursor dose on the growth of 9-AGNRs, their length evolution, quality of alignment, and overall surface disorder,– which are crucial parameters for the growth and transfer of high-quality GNR samples for device integration.

**Methods**

*On-surface synthesis and STM characterization of 9-AGNRs*

Au(788) single crystal growth substrate (MaTecK GmbH, Germany) was cleaned in ultra-high vacuum (UHV) with two cycles of sputtering at 1 kV Ar+ (for 10 minutes) and annealing at 420 °C (for 10 minutes). Subsequently, the 9-AGNR precursor monomer, 3′,6′-di-iodine-1,1′:2′,1″-terphenyl (DITP)[39], was sublimated onto the pristine Au surface from a quartz crucible heated to 70 °C, while the substrate remained at room temperature. To control the deposition rate, a quartz microbalance was employed to maintain a constant deposition rate of 1 Å/min. This deposition rate is not calibrated to accurately correspond to the true surface coverage. Instead, it is calibrated relative to a standard measurement obtained through STM, by counting the number of GNRs. 8-9 Å corresponds to the amount of precursor molecules resulting in GNR monolayer saturation coverage. Following deposition, with deposition times varying to afford from 1 to 9 Å deposits, the substrate was heated to 200 °C (0.5 K/s) for 10 minutes to initiate DITP polymerization, followed by annealing at 400 °C (0.5 K/s) for 10 minutes to induce cyclodehydrogenation. After the sample cooled down to room temperature (RT), STM images were acquired at RT in constant current mode, typically with a -1.5 V sample bias and a 0.03 nA setpoint current using a Scienta Omicron VT-STM.

*Transfer of GNRs to ROS*

To transfer the 9-AGNRs from the Au(788) growth substrate to ROS, electrochemical delamination was used. First, a support layer of poly(methyl methacrylate) (PMMA) was spin-coated (4 PMMA layers, 2500 rpm for 90 s) on the 9-AGNR/Au(788) surface, followed by a 10-minute curing process at 80 °C. PMMA was removed from the Au(788) crystal's edges using a two-step process: 80-minute UV exposure (leading to the breakdown of the chemical bonds in the PMMA), followed by a 3-minute development in water/isopropanol (to remove the PMMA from the surface's edges).

Electrochemical delamination was performed in an aqueous solution of NaOH (1 M) as the electrolyte. A DC voltage of 5V (current ~0.2 A) was applied between the PMMA/9-AGNR/Au(788) cathode and a glassy carbon electrode used as the anode. At the interface between PMMA/GNRs and Au, hydrogen bubbles form, resulting from the water reduction: $2H_2O(l) + 2e^- \rightarrow H_2(g) + 2OH^-(aq)$. The $H_2$ bubbles mechanically delaminate the PMMA/GNR layer from the Au(788) surface. The delaminated PMMA/GNR layer was left in ultra-pure water for 5 minutes before being transferred to the target substrate. The delaminated PMMA/GNR layer was transferred to the ROS followed by a two-step annealing: 80 °C for 10 minutes + 110 °C for 20 minutes to improve adhesion between the ROS and the PMMA/GNR layer. Finally, PMMA was dissolved in acetone for 15 minutes, and the resulting GNR/ROS was rinsed with ethanol and ultra-pure water.

*Raman spectroscopy*

Raman spectroscopy measurements were performed using a WITec confocal Raman microscope (WITec Alpha 300R) with a 785 nm (1.5 eV) laser line and a power of 40 mW. A 50x microscope objective was used to focus the laser beam on the sample and collect the scattered light. The Raman spectra were calibrated using the Si peak at 520.5 $cm^{-1}$. The laser wavelength, power, and integration time were optimized for each substrate to maximize the signal while minimizing sample damage. To prevent sample damage, a Raman mapping approach with a size of 10×10 pixels (10×10 μm) was employed, and the measurements were conducted in a home-built vacuum chamber at a pressure of approximately $10^{-2}$ mbar. Polarized Raman measurements were conducted without the analyzing polarizer to collect all the scattered light. A motorized half-wave plate was used to change the polarization direction of the incident laser beam from -90° to +90° in steps of 10°. The scattered signal, with an excitation wavelength of 785 nm, was detected with a 300 mm lens-based spectrometer equipped with a grating of 300 g $mm^{-1}$ (grooves/mm) and a cooled deep-depletion CCD.

Using the WITec software, a cosmic ray filter was applied to all raw maps to remove signatures of photoluminescence. Afterward, the Raman maps were averaged and polynomial background subtraction was applied, followed by batched fitting with a Lorentzian function for all polarization angles between -90° to 90° for each Raman mode. The fitting using the equations mentioned in the results and discussion was done in IGOR Pro software (Wavemetrics Inc.), and the fitting parameters were obtained through the lowest stable Chi-square values.

**Supporting information**

Additional experimental data on: length evolution of 9-AGNRs on Au(788), average full-width at half maximum, peak positions $P$, $\sigma$, and $OD$ for 9-AGNRs on Au and after transfer across all PDs. A short discussion on the impact of inefficient transfer on GNR quality and alignment.

**Acknowledgements**

This work was supported by the Swiss National Science Foundation under grant no. 200020_182015, the European Union Horizon 2020 research and innovation program under grant agreement no. 881603 (GrapheneFlagship Core 3) the European Union Horizon 2020 FET Open project no. 767187 (QuIET), and the Office of Naval Research BRC Program under the grant N00014-18-1-2708. The authors also greatly appreciate the financial support from the Werner Siemens Foundation (CarboQuant). R.D. acknowledges funding from the University of Bern.

# The role of precursor coverage in the synthesis and substrate transfer of graphene nanoribbons

*Rimah Darawish[1,2], Oliver Braun[3,4], Klaus Müllen[5,6], Michel Calame[3,4,7], Pascal Ruffieux[1], Roman Fasel[1,2] and Gabriela Borin Barin[1\*]*

[1]Empa, Swiss Federal Laboratories for Materials Science and Technology, nanotech@surfaces Laboratory, 8600 Dübendorf, Switzerland

[2]Department of Chemistry, Biochemistry and Pharmaceutical Sciences, University of Bern, 3012 Bern, Switzerland

[3]Empa, Swiss Federal Laboratories for Materials Science and Technology, Transport at Nanoscale Interfaces Laboratory, 8600 Dübendorf, Switzerland

[4]Department of Physics, University of Basel, 4056 Basel, Switzerland

[5]Max Planck Institute for Polymer Research, 55128 Mainz, Germany

[6]Department of Chemistry, Johannes Gutenberg University Mainz, Duesbergweg 10-14, 55128, Mainz, Germany

[7]Swiss Nanoscience Institute, University of Basel, 4056 Basel, Switzerland

*corresponding author: gabriela.borin-barin@empa.ch

**Length evolution of 9-AGNRs on Au(788)**

| Precursor dose | 1Å | 2Å | 3Å | 4Å | 5Å | 6Å | 7Å | 8-9Å |
|---|---|---|---|---|---|---|---|---|
| GNR length at 1st-row position | 14± 2 | 19± 5 | 35± 4 | 39± 4 | 43± 3 | 44± 2 | 45± 2 | 46± 2 |
| GNR length at 2nd-row position | | | | 14± 5 | 22± 3 | 30± 2 | 36± 3 | 42± 5 |
| GNR length at 3rd-row position | | | | | | | 13± 2 | 36±5 |

**Table S1.** Length evolution of 9-AGNRs on Au(788). Average length (in nm) of GNRs at the 1st-, 2nd-, and 3rd-row positions for 30 samples of 9-AGNRs with different PDs ranging from 1-9 Å, along with their respective standard deviation.

**9-AGNR quality and alignment as a function of PD on Au(788) and after substrate transfer**

| Raman mode | RBLM | | CH | | D | | G | |
|---|---|---|---|---|---|---|---|---|
| Substrate | Au(788) | ROS | Au(788) | ROS | Au(788) | ROS | Au(788) | ROS |
| 1 Å | 311±1 | 313±1 | 1242±3 | 1231±1 | 1334±1 | 1337±1 | 1596±1 | 1593±1 |
| 2 Å | 311±1 | 312±1 | 1245±3 | 1235±2 | 1335±2 | 1337±1 | 1596±1 | 1593±2 |
| 3 Å | 312±1 | 313±1 | 1245±2 | 1235±1 | 1336±1 | 1338±1 | 1597±1 | 1593±2 |
| 4 Å | 312±2 | 313±1 | 1242±4 | 1235±1 | 1334±3 | 1337±1 | 1597±1 | 1594±2 |
| 5 Å | 311±1 | 314±1 | 1246±2 | 1234±2 | 1334±1 | 1337±1 | 1596±1 | 1594±1 |
| 6 Å | 311±1 | 313±1 | 1247±2 | 1236±2 | 1336±2 | 1338±1 | 1596±1 | 1594±2 |
| 7 Å | 311±1 | 312±1 | 1247±3 | 1235±2 | 1338±2 | 1339±1 | 1597±1 | 1595±2 |
| 8-9 Å | 312±1 | 313±1 | 1245±2 | 1232±2 | 1339±2 | 1339±1 | 1597±1 | 1594±1 |

**Table S2.** The average peak position (in cm$^{-1}$) of RBLM, CH, D, and G modes measured for PDs from 1-9 Å in samples on Au(788) (40 samples) and after substrate transfer onto ROS (27 samples). Data are obtained from Raman maps acquired in a vacuum chamber using a 785 nm excitation wavelength.

| Raman mode | RBLM | | CH | | D | | G | |
|---|---|---|---|---|---|---|---|---|
| Substrate | Au(788) | ROS | Au(788) | ROS | Au(788) | ROS | Au(788) | ROS |
| 1 Å | 14±1 | 15±2 | 30±5 | 35±4 | 13±2 | 14±1 | 11±2 | 13±1 |
| 2 Å | 13±1 | 16±2 | 30±6 | 38±6 | 14±2 | 15±1 | 12±2 | 12±1 |
| 3 Å | 15±2 | 17±5 | 29±5 | 33±9 | 13±2 | 19±8 | 13±2 | 12±1 |
| 4 Å | 13±1 | 15±3 | 32±6 | 37±4 | 14±2 | 17±5 | 13±2 | 14±2 |
| 5 Å | 14±1 | 16±1 | 32±5 | 33±4 | 13±1 | 16±4 | 12±2 | 14±1 |
| 6 Å | 14±1 | 16±1 | 32±5 | 33±4 | 15±1 | 16±4 | 13±1 | 14±1 |
| 7 Å | 16±1 | 15±1 | 33±5 | 32±4 | 17±2 | 18±4 | 14±2 | 15±2 |
| 8-9 Å | 16±1 | 15±2 | 32±4 | 32±4 | 17±2 | 18±4 | 14±2 | 15±2 |

**Table S3**. The average full-width at half maximum (FWHM, in cm$^{-1}$) of RBLM, CH, D, and G modes measured for PDs from 1-9 Å in samples on Au(788) (40 samples) and after substrate transfer onto ROS (27 samples). Data are obtained from Raman maps acquired in vacuum using a 785 nm excitation wavelength.

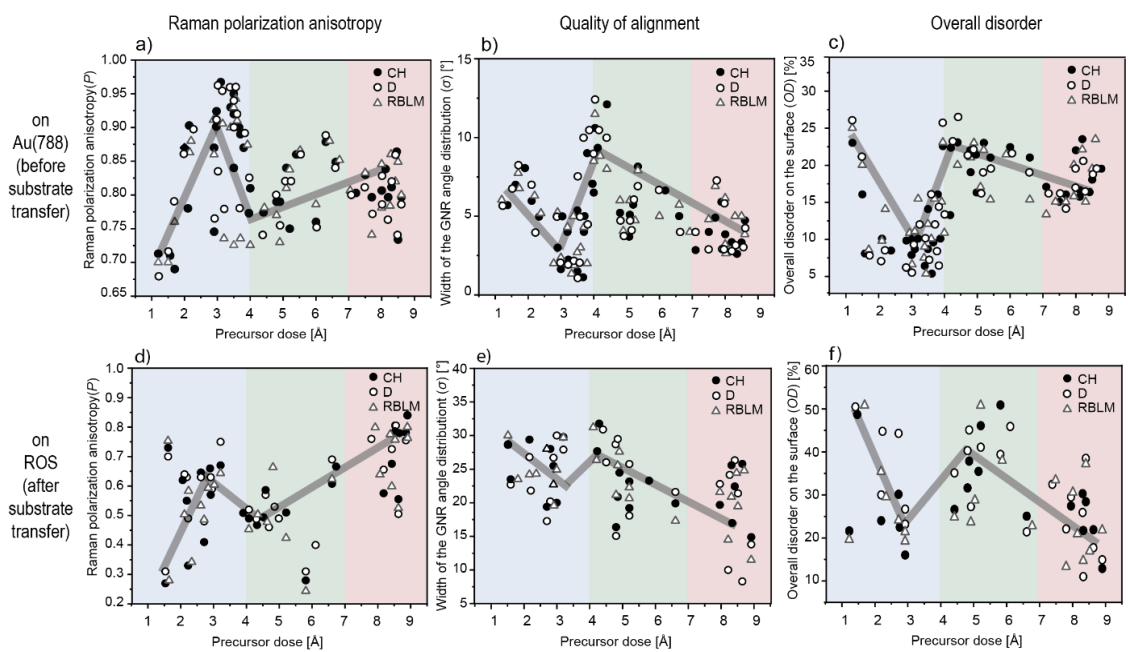

**Figure S1: Impact of PD on the alignment of 9-AGNRs on Au(788) and after substrate transfer onto ROS**. The alignment was evaluated using Eq (2) from polarized Raman data for RBLM, CH, and D modes. Panels (a), (b), and (c) show the average $P$, $\sigma$, and $OD$ of RBLM, CH, and D modes, respectively, as a function of PD from 1-9 Å on Au(788) and panels (d), (e) and (f) after substrate transfer onto ROS. The colored background areas represent the growth regimes in which GNRs predominantly grow at the 1$^{st}$-row position (blue), at the 2$^{nd}$-row position (green), and at the 3$^{rd}$-row position (red). The data are acquired in a vacuum using 785 nm excitation.

**The impact of inefficient substrate transfer on GNR integrity and alignment**

Here, we investigate the effect of inefficient substrate transfer on the integrity and alignment of GNRs in low-PD samples (3 Å) using Raman characterization. Figure S2 shows the Raman spectra and polar plot for a sample with PD = 3 Å on the growth surface Au(788) and after inefficient substrate transfer onto ROS. Table S4 presents the FWHM and peak position for the RBLM, CH, D, and G modes acquired with the 9-AGNRs on Au(788) and after their transfer onto ROS. Our observations indicate that inefficient substrate transfer resulted in an increase in FWHM and peak shifts for all active Raman modes. Specifically, we observe a large shift of -10 cm$^{-1}$ for the RBLM and a significant shift of +44 cm$^{-1}$ for the CH mode. The FWHM of RBLM, CH, D, and G modes broaden from 11, 31, 12, and 13 cm$^{-1}$ on Au(788) to 25, 100, 85, and 44 cm$^{-1}$ on ROS, respectively, confirming the inefficiency of the substrate transfer for this sample. Changes in the CH-D region can be attributed to damage in the GNR edge structure[1–3], while changes in the G and RBLM modes suggest the presence of defects and/or doping[4–6].

To investigate the influence of inefficient substrate transfer on the alignment and overall disorder on the surface (*OD*), we extract *P*, *σ*, and *OD* on both substrates for the G mode (Table S5). We observe a significant decrease in *P* along with broadening of *σ*: $P_{Au(788)}$= 0.92 ( $\sigma_{Au(788)}$= 1°) to $P_{ROS}$= 0.40 ($\sigma_{ROS}$= 40°). Additionally, *OD* increases from 14% on Au (788) to 35% on ROS. These findings confirm that GNRs are susceptible to defects and can vary in quality upon electrochemical delamination transfer.

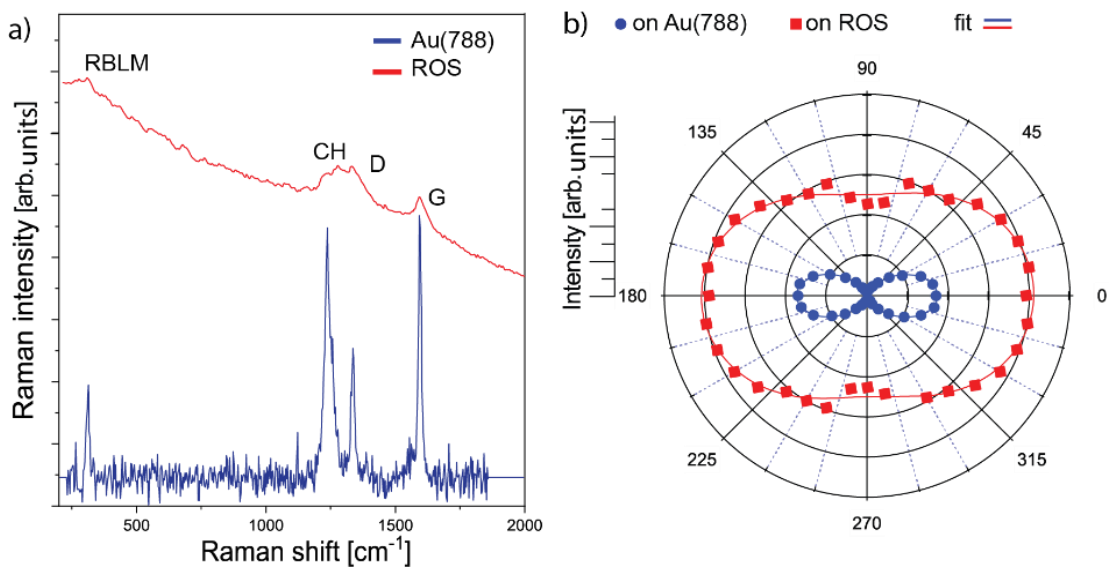

**Figure S2.** Raman spectra and G-mode polar plots for a 9-AGNR sample grown with PD = 3 Å on Au(788) and after inefficient substrate transfer onto the ROS. (a) Raman spectra on Au(788) in blue and on the ROS in red. (b) Polar plot of the Raman intensity of the G-mode on Au(788) in blue and on the ROS in red. Spectra were acquired using an excitation wavelength of 785 nm under vacuum conditions and changing the incident laser's angle ($\theta_{in}$) with respect to GNRs' alignment direction ($\theta = 0°$), without utilizing a polarizer in the detection path.

**Table S4.** Peak position and FWHM of the RBLM, CH, D, and G modes for the sample in Fig. S2 on Au(788), and after transfer onto ROS.

| Substrate | Raman mode | RBLM | CH | D | G |
|---|---|---|---|---|---|
| Au(788) | Peak position [cm$^{-1}$] | 312 | 1240 | 1339 | 1593 |
| | FWHM [cm$^{-1}$] | 11 | 31 | 12 | 13 |
| ROS | Peak position [cm$^{-1}$] | 302 | 1284 | 1337 | 1591 |
| | FWHM [cm$^{-1}$] | 25 | 100 | 85 | 44 |

**Table S5.** *P*, *σ*, and *OD* of the G mode for the sample in Fig. S2 on Au(788), and after transfer onto ROS.

| Substrate | Au(788) | ROS |
|---|---|---|
| Raman polarization anisotropy (*P*) | 0.92 | 0.4 |
| Quality of alignment (*σ*) [°] | 1±0.05 | 40±0.8 |
| Overall disorder on the surface (*OD*) [%] | 14 | 35 |